\documentclass[showpacs,fleqn,nobibnotes,twocolumn]{revtex4}

\usepackage{amsmath}
\usepackage{graphicx}

\def\lsim{\raise0.3ex\hbox{$<$\kern-0.75em\raise-1.1ex\hbox{$\sim$}}}
\def\gsim{\raise0.3ex\hbox{$>$\kern-0.75em\raise-1.1ex\hbox{$\sim$}}}

\newcommand{\rr}{\mbox{\boldmath $r$}}

\newcommand{\rb}{\mbox{\boldmath $b$}}

\newcommand{\rd}{\mbox{\boldmath $\Delta$}}

\newcommand{\N}{\mathcal{N}}

\begin{document}

\title{Investigating the impact of the gluon saturation effects on the momentum transfer distributions for the exclusive vector meson photoproduction in hadronic collisions}
\author{V. P. Gon\c{c}alves $^{1}$, F. S. Navarra $^{2}$  and D. Spiering $^{2}$}

\affiliation{$^{1}$ Instituto de F\'{\i}sica e Matem\'atica,  Universidade
Federal de Pelotas, 
Caixa Postal 354, CEP 96010-900, Pelotas, RS, Brazil}
\affiliation{$^{2}$ Instituto de F\'{\i}sica, Universidade de S\~{a}o Paulo, CEP 05315-970 S\~{a}o Paulo, SP, Brazil.}

\begin{abstract}
The exclusive vector meson production cross section is one of the most promising observables to probe 
the high energy regime of the QCD dynamics. In particular, 
the squared  momentum transfer ($t$) distributions are an important source of information about the 
spatial distribution of the gluons in the hadron and about fluctuations  of the  color fields. In this 
paper we complement previous studies on  exclusive vector meson photoproduction in hadronic collisions 
presenting a  comprehensive analysis of the $t$ - spectrum measured in exclusive $\rho$, $\phi$ and 
$J/\Psi$ photoproduction in $pp$ and $PbPb$ collisions at the LHC. We compute the  differential cross 
sections  taking into account  gluon saturation  effects and compare the predictions with those obtained 
in the linear regime of the QCD dynamics. 
Our results show that  gluon saturation  suppresses the magnitude of the cross sections and shifts the 
position of the dips towards smaller values of $t$. 

\end{abstract}

\maketitle

\section{Introduction}

Experimental results released in the last years have demonstrated that photon -- induced interactions 
in hadronic collisions can be used to probe several aspects of the Standard Model (SM) as well as to test 
predictions of  Beyond SM Physics (For a recent review see Ref. \cite{review_forward}). In particular, 
the study of the exclusive vector meson photoproduction in hadronic collisions is an important source of information about the hadronic structure and also about QCD dynamics at high energies \cite{vicbert,vicmag}. 
As exclusive processes are driven by the gluon  content of the target, with the  cross sections being  proportional to the square of the scattering amplitude,  they  are strongly sensitive to the underlying 
QCD dynamics.   Additionally, the squared  momentum transfer ($t$) distributions give access to the spatial distribution of the gluons in the hadron and about fluctuations  of the  color fields (See e.g. Ref. \cite{heike}).

In the last years  exclusive vector meson photoproduction in hadronic collisions has
been discussed by several authors considering different assumptions and distinct approaches (See e.g. Refs. \cite{vicmag_varios,schafer,jones,guzey,bruno,magno}). 
In particular, in Refs. \cite{bruno,bruno_run2} we demonstrated that the experimental LHC Run 1 data and the preliminary Run 2 data can be sucessfully described within the color dipole formalism if  non - linear 
effects in the QCD dynamics are taken into account. The main advantage of this approach is that the 
main ingredients can be constrained by the very precise HERA data and hence the predictions for photon -- 
induced interactions at the LHC are parameter free. 
In those previous works we presented our predictions for the $t$ -- integrated observables -- rapidity distributions and total cross sections -- which have been measured by the ALICE, CMS and LHCb Collaborations at the LHC in the Run 1. In principle, the $t$ - distributions may be measured in Run 2 \cite{review_forward}. This encourages
us to  extend our previous studies and present the color dipole predictions for the 
$t$ -- spectrum measured in exclusive vector meson photoproduction in hadronic collisions. In particular, 
in this paper we will use the  color dipole formalism to describe the photon - hadron interaction, with the scattering amplitude being expressed in terms  of the impact parameter Color Glass Condensate  (bCGC) model, which successfully describes the $t$ - distributions for the exclusive vector meson production at HERA. We will compute the $t$ - spectrum for the exclusive $\rho$, $\phi$ and $J/\Psi$ photoproduction in $pp$ and $PbPb$ collisions at the LHC energies probed in the Run 2. Moreover, in the case of $PbPb$ collisions, we will consider the coherent and incoherent contributions to  exclusive production, which are associated to processes where the nucleus target scatters elastically or breaks up, respectively. For a similar analysis considering alternative approaches see Ref. \cite{guzey_tdist}. In order to investigate the impact of the  gluon saturation effects, associated to non - linear contributions for the QCD dynamics at high energies, we will compare our predictions with those obtained disregarding these effects, i. e. using a linear model for the QCD dynamics. 
As the dipole formalism of exclusive processes has been discussed in detail in our previous works 
\cite{bruno,bruno_run2,diego1,diego2}, in the next Section we will only review the main elements needed 
to study exclusive vector meson photoproduction in hadronic collisions. In Section \ref{res} we will present our predictions for the rapidity and  $t$ -- distributions and in Section \ref{conc} we will summarize our main conclusions.

\section{Formalism}
An ultra relativistic charged hadron (proton or nucleus) 
gives rise to strong electromagnetic fields. In a hadronic collision, the photon stemming from the 
electromagnetic field of one of the two colliding hadrons can 
interact with one photon of the other hadron (photon - photon process) or can interact directly with the other hadron (photon - hadron process) \cite{upc}. In the particular case of  exclusive vector meson 
photoproduction in hadronic collisions, the differential cross section can be expressed as follows
\begin{widetext}
\begin{eqnarray}
\frac{d\sigma \,\left[h_1 + h_2 \rightarrow   h_1 \otimes V \otimes h_2\right]}{dY\,dt}  =  \left[\omega \frac{dN}{d\omega}|_{h_1}\,\frac{d\sigma}{dt}(\gamma h_2 
\rightarrow V \otimes h_2)\right]_{\omega_L} 
 +  \left[\omega \frac{dN}{d\omega}|_{h_2}\,\frac{d\sigma}{dt}(\gamma h_1 \rightarrow V \otimes h_1)\right]_{\omega_R}\,
\label{dsigdy}
\end{eqnarray}
\end{widetext}
where the rapidity ($Y$) of the vector meson in the final state is determined by the photon energy $\omega$ in the collider frame and by the mass $M_{V}$ 
of the vector meson [$ Y \propto \ln \, ( \omega/M_{V})$]. Moreover, $d\sigma/dt$ is the differential cross section for the $\gamma h_i \rightarrow V \otimes h_i$ process, with the symbol $\otimes$ representing the presence of a rapidity gap in the final state and $\omega_L \, (\propto e^{-Y})$ and $\omega_R \, (\propto e^{Y})$ denoting  photon energies from the $h_1$ and $h_2$ hadrons, respectively. Furthermore, $\frac{dN}{d\omega}$ denotes the  equivalent photon 
spectrum  of the relativistic incident hadron, with the flux of a nucleus   being enhanced by a factor $Z^2$ in comparison to the proton one. 
Eq. (\ref{dsigdy}) takes into account the fact that both incident hadrons can be sources of the photons which will interact with the other hadron, with 
the first term on the right-hand side of the Eq. (\ref{dsigdy}) being dominant  at positive rapidities while the second term dominating at negative 
rapidities due to the fact that the photon flux has support at small values of $\omega$, decreasing exponentially at large $\omega$. As in Refs.  
\cite{bruno,bruno_run2} we will assume that the photon flux associated to the proton and to the  nucleus 
can be described by  the Dress - Zeppenfeld  \cite{Dress} and the relativistic point -- like 
charge \cite{upc} models, respectively.

In the color dipole formalism, the $\gamma h \rightarrow V h$ process can be factorized in terms of the fluctuation of the virtual photon into a $q \bar{q}$ color dipole, the dipole-hadron scattering by a color singlet exchange  and the recombination into the vector meson  $V$. The final state is characterized by the presence of a rapidity gap. The differential cross section for the exclusive vector meson photoproduction can be expressed as follows
\begin{eqnarray}
\frac{d\sigma}{dt}
& = & \frac{1}{16\pi}  |{\cal{A}}^{\gamma h \rightarrow V h }(x,  \Delta)|^2\,\,,
\label{dsigdt}
\end{eqnarray}
with the  amplitude for producing an exclusive vector meson diffractively  being given in the color dipole formalism by
\begin{widetext}
\begin{eqnarray}
 {\cal A}^{\gamma h \rightarrow V h }({x},\Delta)  =  i
\int dz \, d^2\rr \, d^2\rb_h \,  e^{-i[\rb_h-(1-z)\rr].\rd}  
 \,\, (\Psi^{V*}\Psi)  \,\,2 {\cal{N}}^h({x},\rr,\rb_h)
\label{amp}
\end{eqnarray}
\end{widetext}
where  $(\Psi^{V*}\Psi)$ denotes the wave function overlap between the  photon and vector meson wave functions, $\Delta = - \sqrt{t}$ is the momentum 
transfer and $\rb_h$ is the impact parameter of the dipole relative to the hadron target. Moreover, the variables  $\rr$ and $z$ are the dipole transverse 
radius and the momentum fraction of the photon carried by a quark (an antiquark carries then $1-z$), respectively. $ {\cal N}^h (x, \rr, \rb_h)$ is the 
forward dipole-target scattering amplitude (for a dipole at  impact parameter $\rb_h$) which encodes all the information about the hadronic scattering, 
and thus about the non-linear and quantum effects in the hadron wave function. It  depends on the $\gamma h$  center - of - mass reaction energy, 
$W = [2 \omega \sqrt{s}]^{1/2}$, through the variable $ x = M^2_V/W^2$. As in Refs. \cite{bruno,bruno_run2}, in what follows we will consider the Boosted 
Gaussian model \cite{KT,KMW} for the overlap function and the impact parameter Color Glass Condensate (bCGC) model \cite{KMW} for the dipole -- proton 
scattering amplitude ${\cal{N}}^p$. 
In this model the dipole - proton scattering amplitude is given by \cite{KMW} 
\begin{widetext}
\begin{eqnarray}
\mathcal{N}^p(x,\rr,\rb_p) =   
\left\{ \begin{array}{ll} 
{\mathcal N}_0\, \left(\frac{ r \, Q_s(b_p)}{2}\right)^{2\left(\gamma_s + 
\frac{\ln (2/r \, Q_s(b_p))}{\kappa \,\lambda \,Y}\right)}  & \mbox{$r Q_s(b_p) \le 2$} \\
 1 - e^{-A\,\ln^2\,(B \, r \, Q_s(b_p))}   & \mbox{$r Q_s(b_p)  > 2$} 
\end{array} \right.
\label{eq:bcgc}
\end{eqnarray}
\end{widetext} 
with  $\kappa = \chi''(\gamma_s)/\chi'(\gamma_s)$, where $\chi$ is the 
LO BFKL characteristic function.  The coefficients $A$ and $B$  
are determined uniquely from the condition that $\mathcal{N}^p(x,\rr,\rb_p)$, and its derivative 
with respect to $r\,Q_s(b_p)$, are continuous at $r\,Q_s(b_p)=2$. The impact parameter dependence of the  proton saturation scale $Q_s(b_p)$  is given by:
\begin{equation} 
  Q_s(b_p)\equiv Q_s(x,b_p)=\left(\frac{x_0}{x}\right)^{\frac{\lambda}{2}}\;
\left[\exp\left(-\frac{{b_p}^2}{2B_{\rm CGC}}\right)\right]^{\frac{1}{2\gamma_s}},
\label{newqs}
\end{equation}
with the parameter $B_{\rm CGC}$  being obtained by a fit of the $t$-dependence of 
exclusive $J/\psi$ photoproduction. The  factors $\mathcal{N}_0$ and  $\gamma_s$  were  
taken  to be free. In what follows we consider the set of parameters obtained in 
Ref. \cite{amir} by fitting the recent HERA data on the reduced $ep$ cross sections:
 $\gamma_s = 0.6599$, $\kappa = 9.9$, $B_{CGC} = 5.5$ GeV$^{-2}$, $\mathcal{N}_0 = 0.3358$, $x_0 = 0.00105$ and $\lambda = 0.2063$.
As demonstrated in Ref. \cite{amir},  these models allow us to successfully describe  the high precision combined 
HERA data on inclusive and exclusive processes.

In  the case of a nuclear target, the exclusive vector meson photoproduction can occur in coherent or incoherent interactions. If the nucleus scatters elastically,  the process is called coherent production. On the other hand, if the nucleus scatters inelastically, i.e. breaks up,   the process is denoted incoherent production. As discussed e.g. in Refs. \cite{Caldwell,Lappi_inc,Toll}, these different processes probe distinct properties of the gluon density of the nucleus. While coherent processes  probe the  average spatial distribution of gluons, the incoherent ones are determined by fluctuations and correlations in the gluon density. As demonstrated e.g. in Refs. \cite{diego1,diego2}, the incoherent processes dominate at large - $t$, with the coherent one being dominant when $t \rightarrow 0$. 
The coherent cross section is given by Eq. (\ref{dsigdt}) in terms of the dipole - nucleus scattering amplitude ${\cal{N}}^A$.  As in our previous works \cite{bruno,bruno_run2,diego1,diego2}, we  will assume that ${\cal{N}}^A$ can be expressed as follows
\begin{eqnarray}
{\cal{N}}^A(x,\rr,\rb_A) = 1 - \exp \left[-\frac{1}{2}  \, \sigma_{dp}(x,\rr^2) 
\,A\,T_A(\rb_A)\right] \,\,,
\label{enenuc}
\end{eqnarray}
where  $T_A(\rb_A)$ is  the nuclear profile function, which is obtained from a 3-parameter Fermi 
distribution form of the nuclear
density normalized to $1$, and  $\sigma_{dp}$ is the dipole-proton cross section that is expressed by
\begin{eqnarray}
\sigma_{dp} = 2 \,\int d^2\rb_p \, {\cal{N}}^p({x},\rr,\rb_p)
\label{sigdip}
\end{eqnarray}
with ${\cal{N}}^p$ given by the bCGC model.
For the calculation of the differential cross section $d\sigma/dt$ 
for incoherent interactions we apply for the vector meson photoproduction the treatment presented in Ref. \cite{Lappi_inc}, which is valid for $t\neq 0$. Consequently, 
we have that 
\begin{widetext}
\begin{eqnarray}
 \frac{d\sigma_{inc}}{dt} = \frac{1}{16\pi}  \int dz dz^{\prime} d^2\rr d^2\rr^{\prime}
 (\Psi^{V*}\Psi)(z,\rr) (\Psi^{V*}\Psi)(z^{\prime},\rr^{\prime}) \, \langle |{\cal{A}}|^2\rangle \,\,,
 \label{difinc}
\end{eqnarray} 
\end{widetext}
 with the average of the squared scattering amplitude  being  approximated by \cite{Lappi_inc}
\begin{widetext}
\begin{eqnarray}
\langle |{\cal{A}}(\rr,\rr^{\prime},t)|^2\rangle = 16\pi^2 B_p^2 &\displaystyle\int& d^2 \rb_A 
\, e^{-B_p\Delta_A^2}\N^p(x,\rr)\N^p(x,\rr^{\prime})\,A\,T_A(\rb_A) \nonumber\\
  &\times&\exp\Big\{ -2\pi(A-1)B_pT_A(\rb_A)\big[\N^p(x,\rr)+\N^p(x,\rr^{\prime})\big] \Big\},
  \label{amp_inc}
\end{eqnarray}
\end{widetext}
where  $\N^p(x,\rr)$ is the dipole - proton scattering amplitude. The parameter $B_p$ is associated to the impact parameter profile function of the proton.

In order to investigate the impact of  gluon saturation effects on the exclusive vector meson 
photoproduction  we will also  estimate the differential cross sections assuming that 
 $\mathcal{N}^p(x,\rr,\rb_p)$ is given by the linear part of the bCGC model, which is 
\begin{eqnarray}
\mathcal{N}^p(x,\rr,\rb_p) =  
{\mathcal N}_0\, \left(\frac{ r \, Q_s(b_p)}{2}\right)^{2\left(\gamma_s + 
\frac{\ln (2/r \, Q_s(b_p))}{\kappa \,\lambda \,Y}\right)}\,,
\label{eq:bcgclin}
\end{eqnarray}
with the same parameters used before in Eq. (\ref{eq:bcgc}).
Moreover, in the case of $\gamma Pb$ interactions we will assume that the  dipole - nucleus amplitude can be expressed  by
\begin{eqnarray}
  \N^A(x,r,\rb_A) =  \frac{1}{2}\sigma_{dp}(x,r)AT_A(\rb_A)
  \label{nalin}
\end{eqnarray}
with $\sigma_{dp}$ expressed by Eq. (\ref{sigdip}) and  $\mathcal{N}^p$ given by Eq. (\ref{eq:bcgclin}). 
Using Eq. (\ref{eq:bcgclin})  we  disregard  possible  non-linear 
effects in the nucleon. On the other hand, using Eq. (\ref{nalin}) we  disregard the multiple 
scatterings of the dipole with the nucleus, which generate non-linear effects in the full calculation.

\begin{figure}[t]
\begin{center}
\scalebox{0.28}{\includegraphics{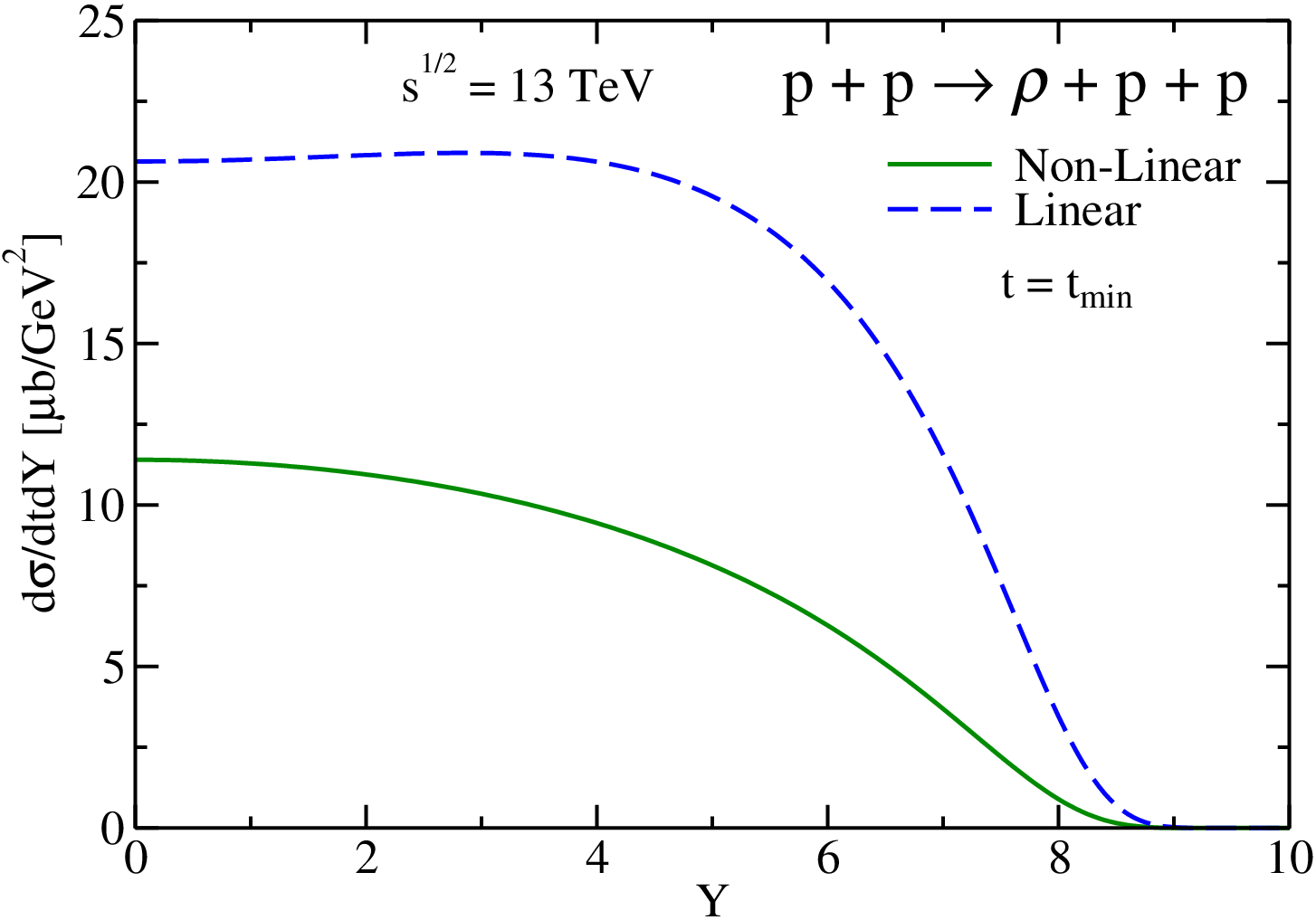}}
\scalebox{0.28}{\includegraphics{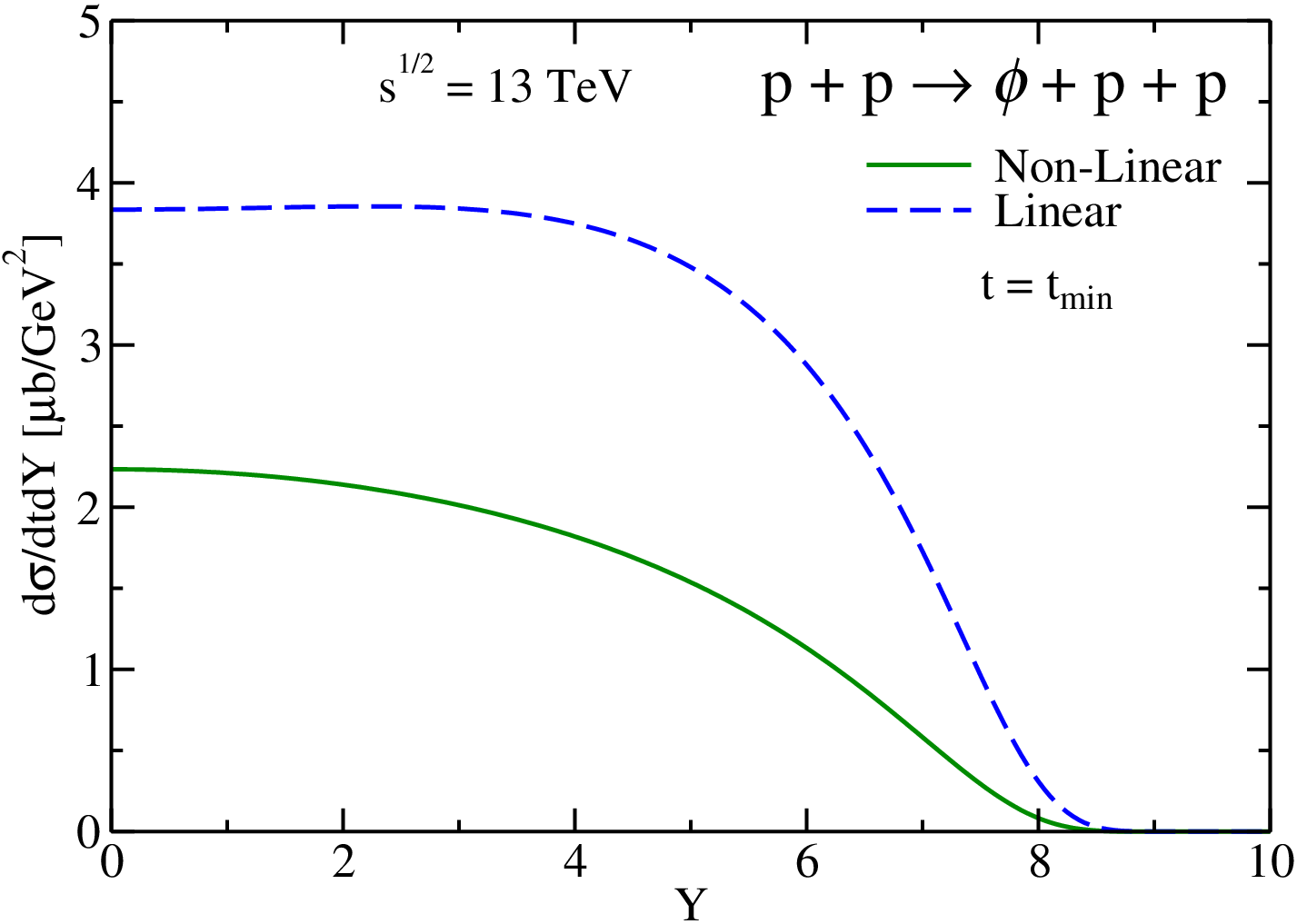}}
\scalebox{0.28}{\includegraphics{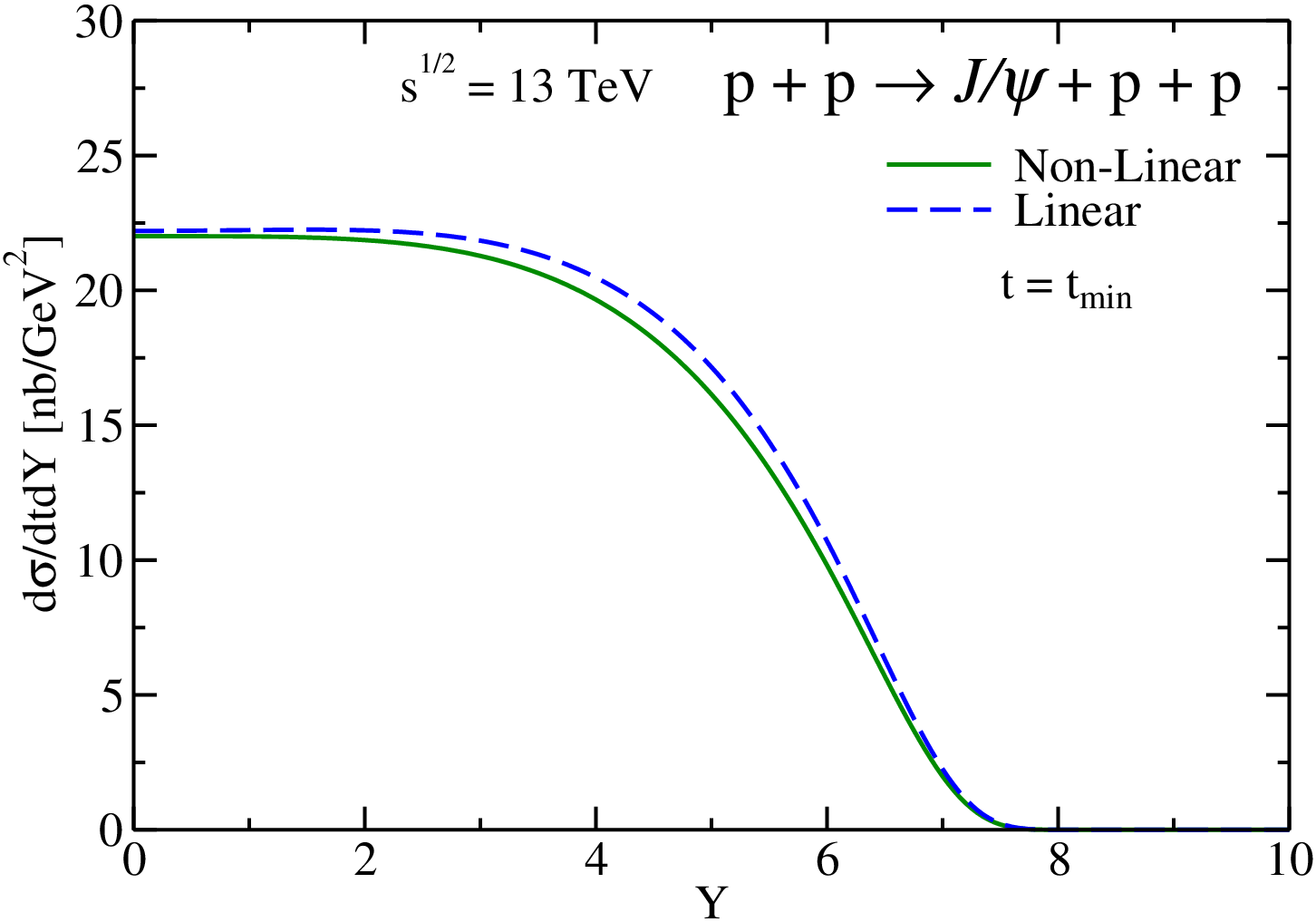}}
\caption{Rapidity distribution for the exclusive $\rho$, $\phi$ and $J/\Psi$ photoproduction in $pp$ collisions at $\sqrt{s} = 13$ TeV.}
\label{Fig:rap_pp}
\end{center}
\end{figure}

\begin{figure}[t]
\begin{center}
\scalebox{0.28}{\includegraphics{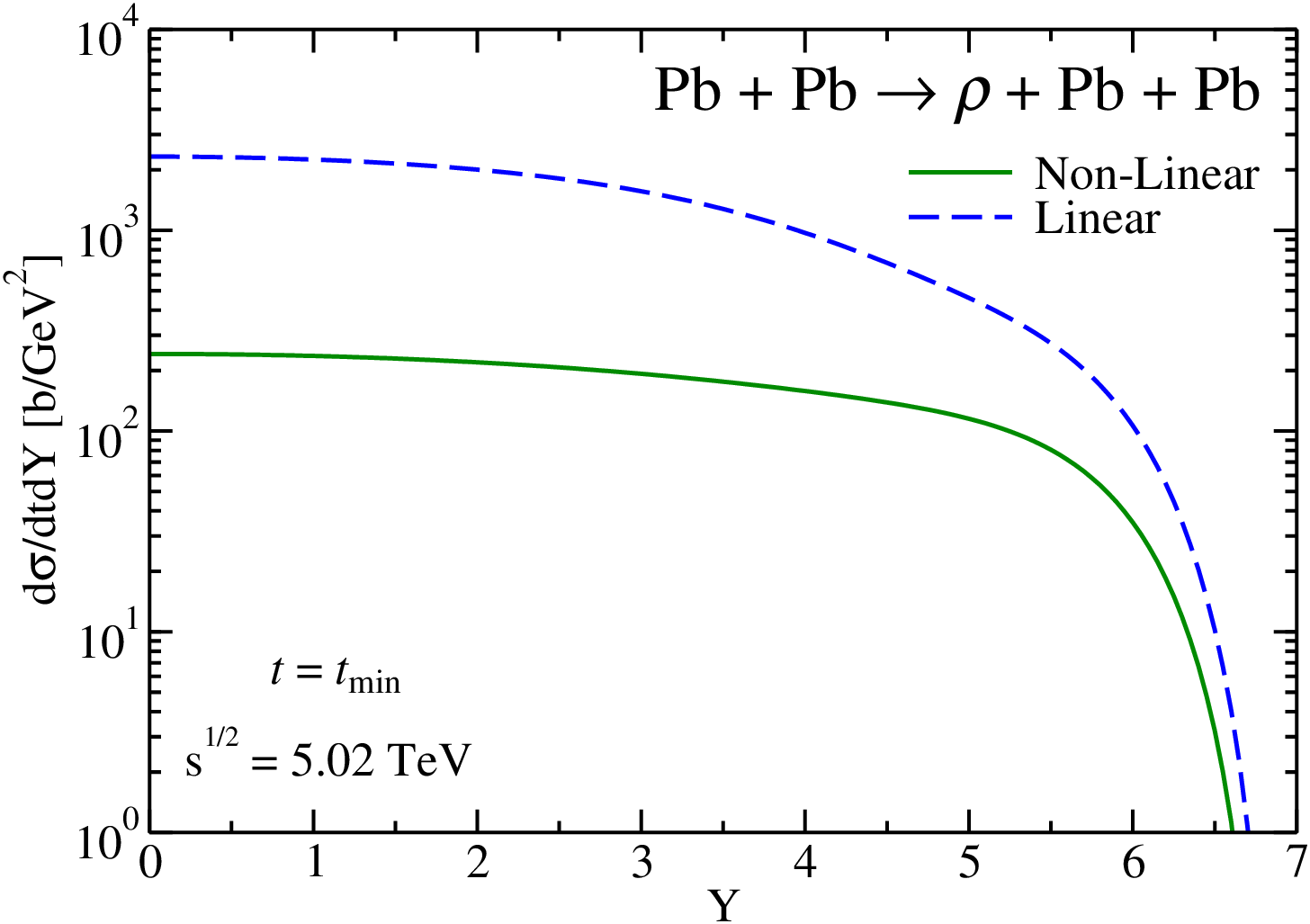}}
\scalebox{0.28}{\includegraphics{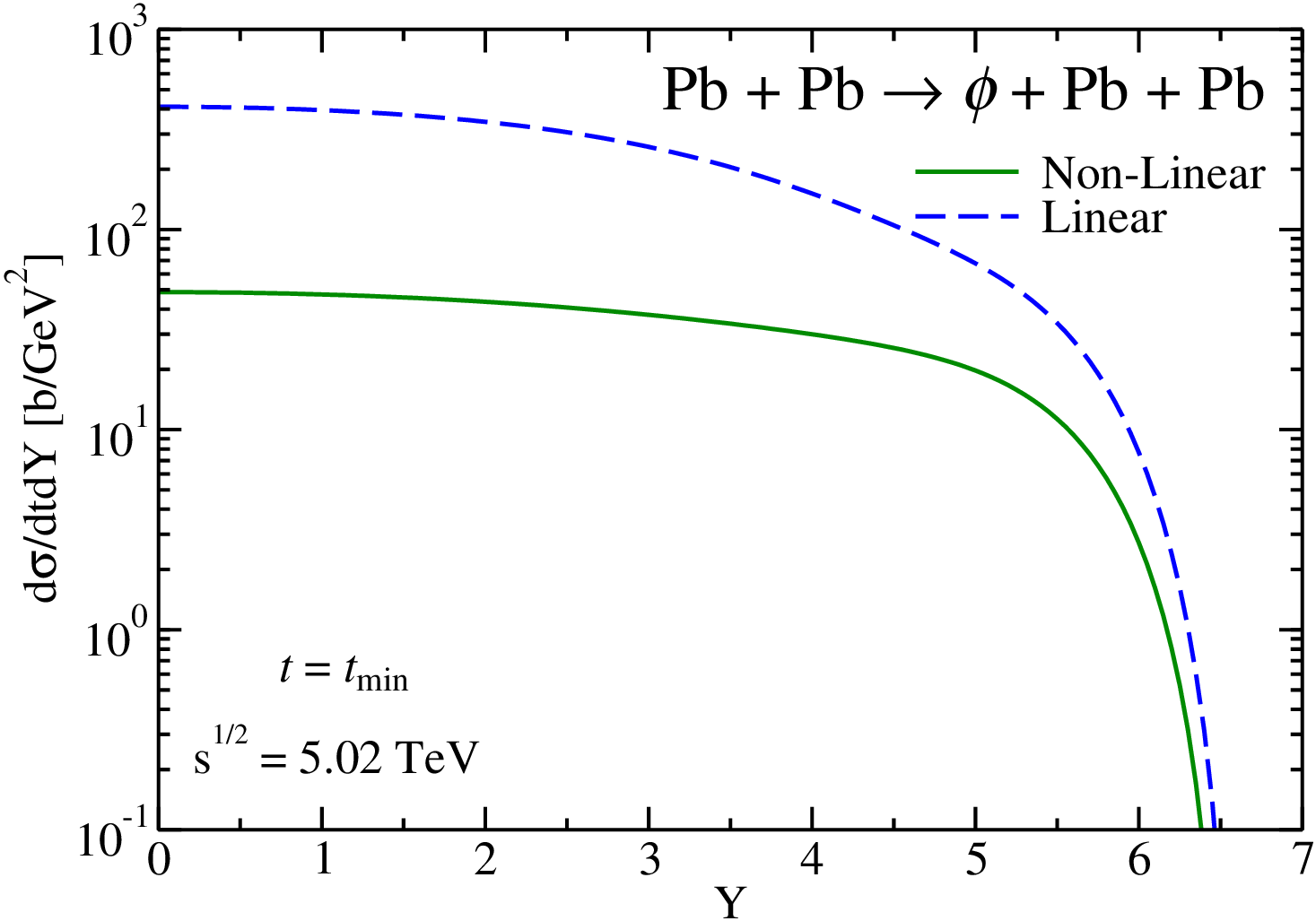}}
\scalebox{0.28}{\includegraphics{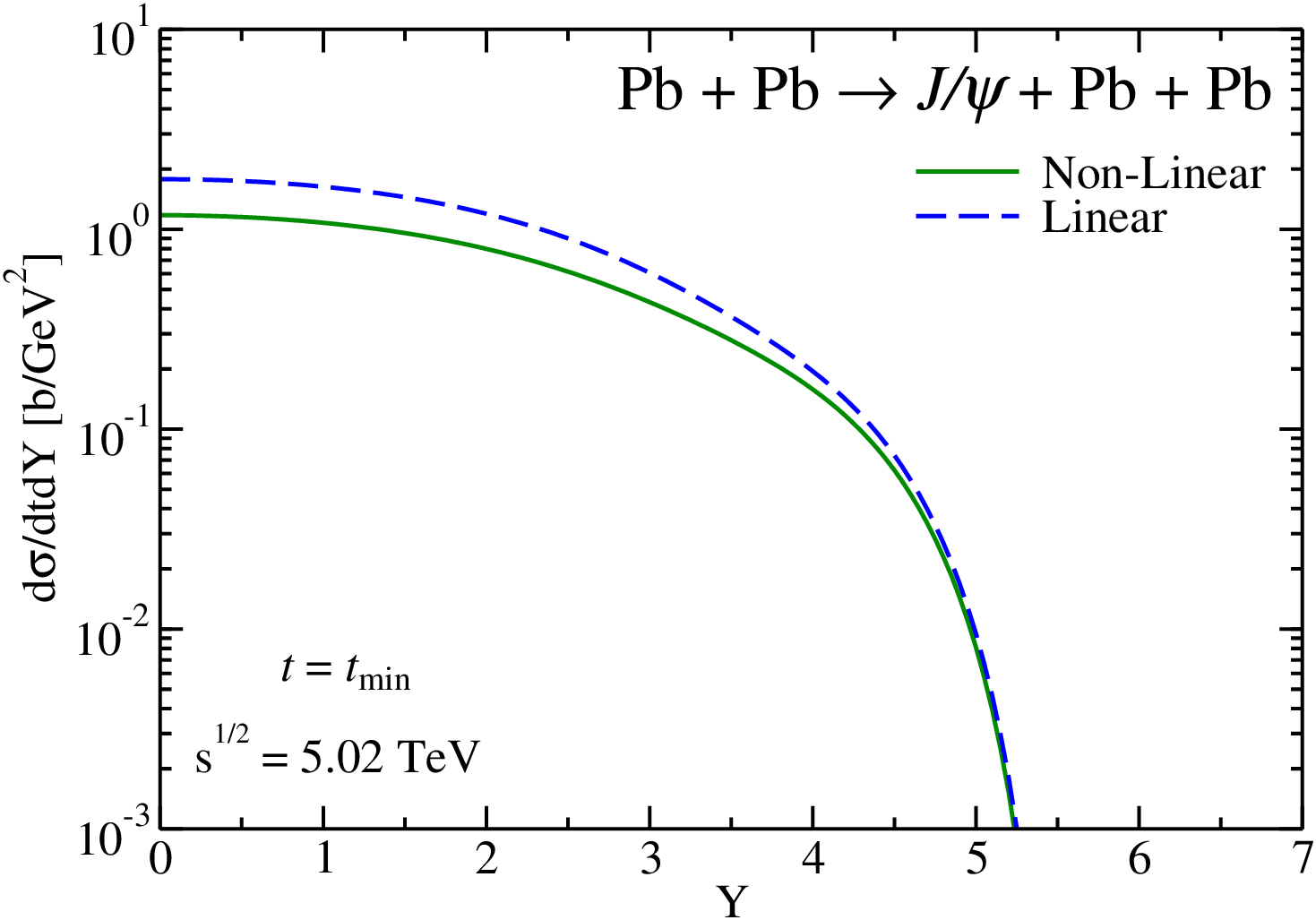}}
\caption{Rapidity distribution for the exclusive $\rho$, $\phi$ and $J/\Psi$ photoproduction in $PbPb$ collisions at $\sqrt{s} = 5.02$ TeV.}
\label{Fig:rap_AA}
\end{center}
\end{figure}

\begin{figure}[t]
\begin{center}
\scalebox{0.28}{\includegraphics{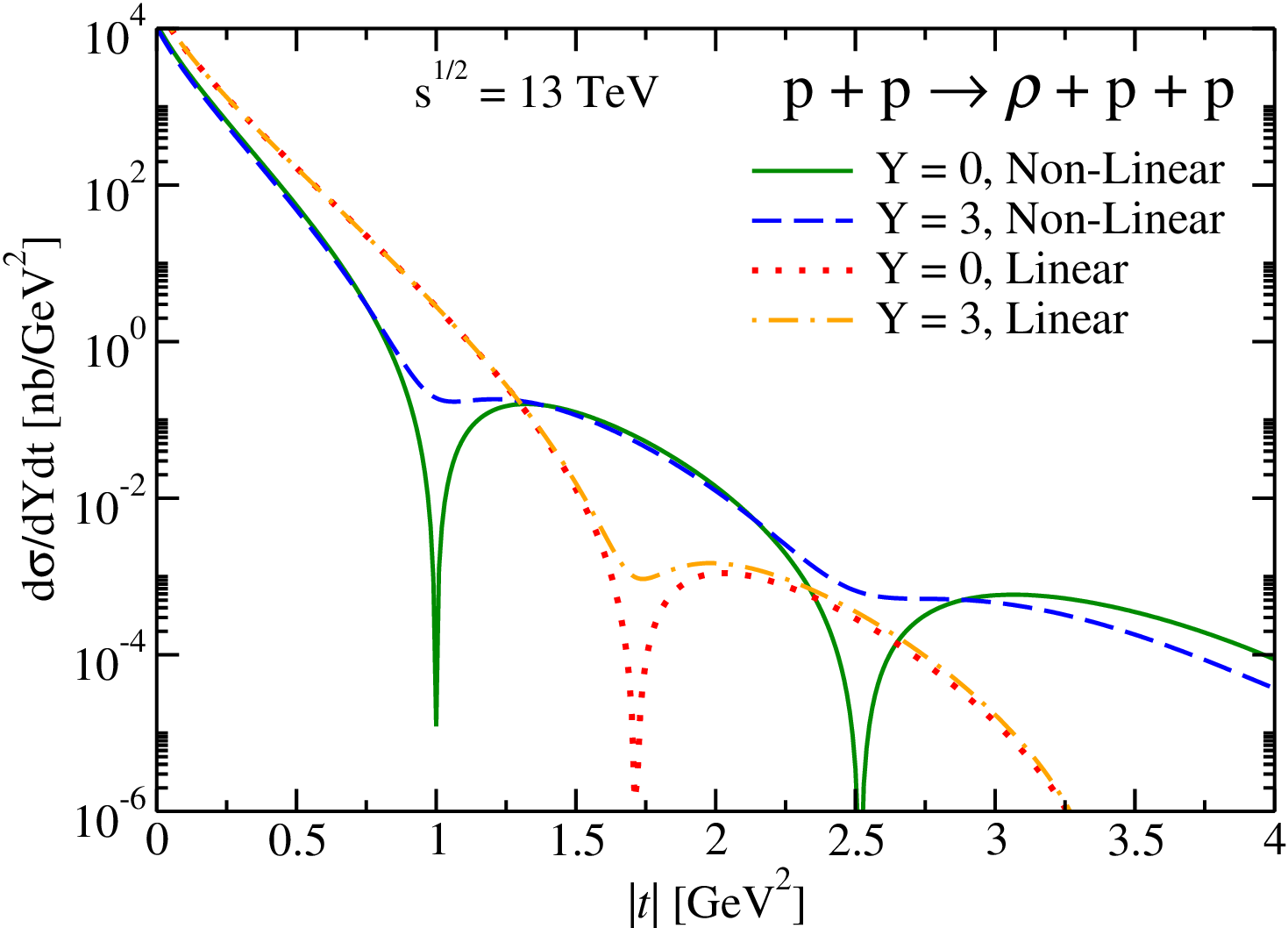}}
\scalebox{0.28}{\includegraphics{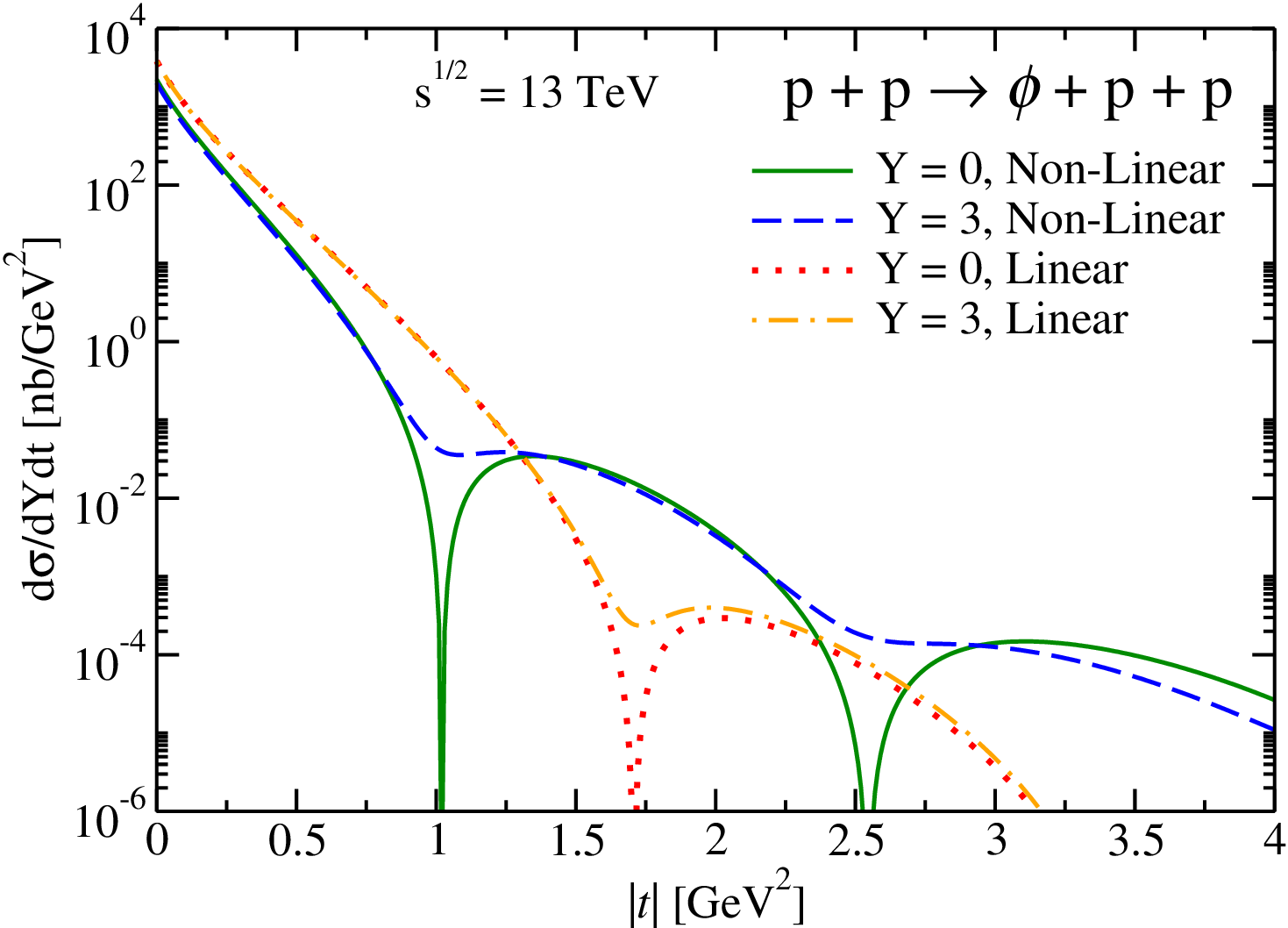}}
\scalebox{0.28}{\includegraphics{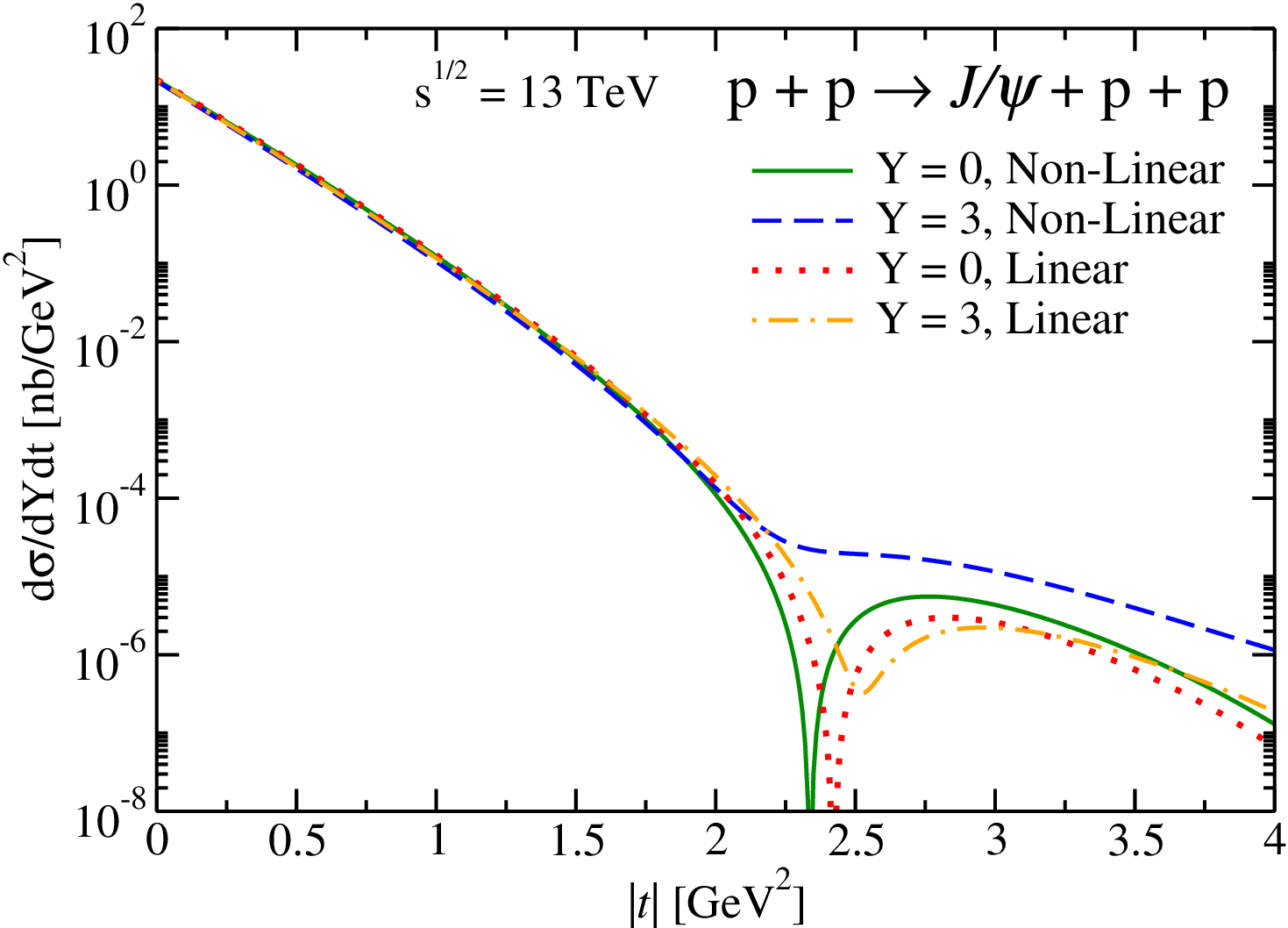}}
\caption{Transverse momentum distributions for the exclusive $\rho$, $\phi$ and $J/\Psi$ photoproduction in $pp$ collisions at $\sqrt{s} = 13$ TeV assuming two different values for the  vector meson rapidity.}
\label{Fig:dsigdt_pp}
\end{center}
\end{figure}

\begin{figure}[!h]
  \includegraphics[scale=0.28]{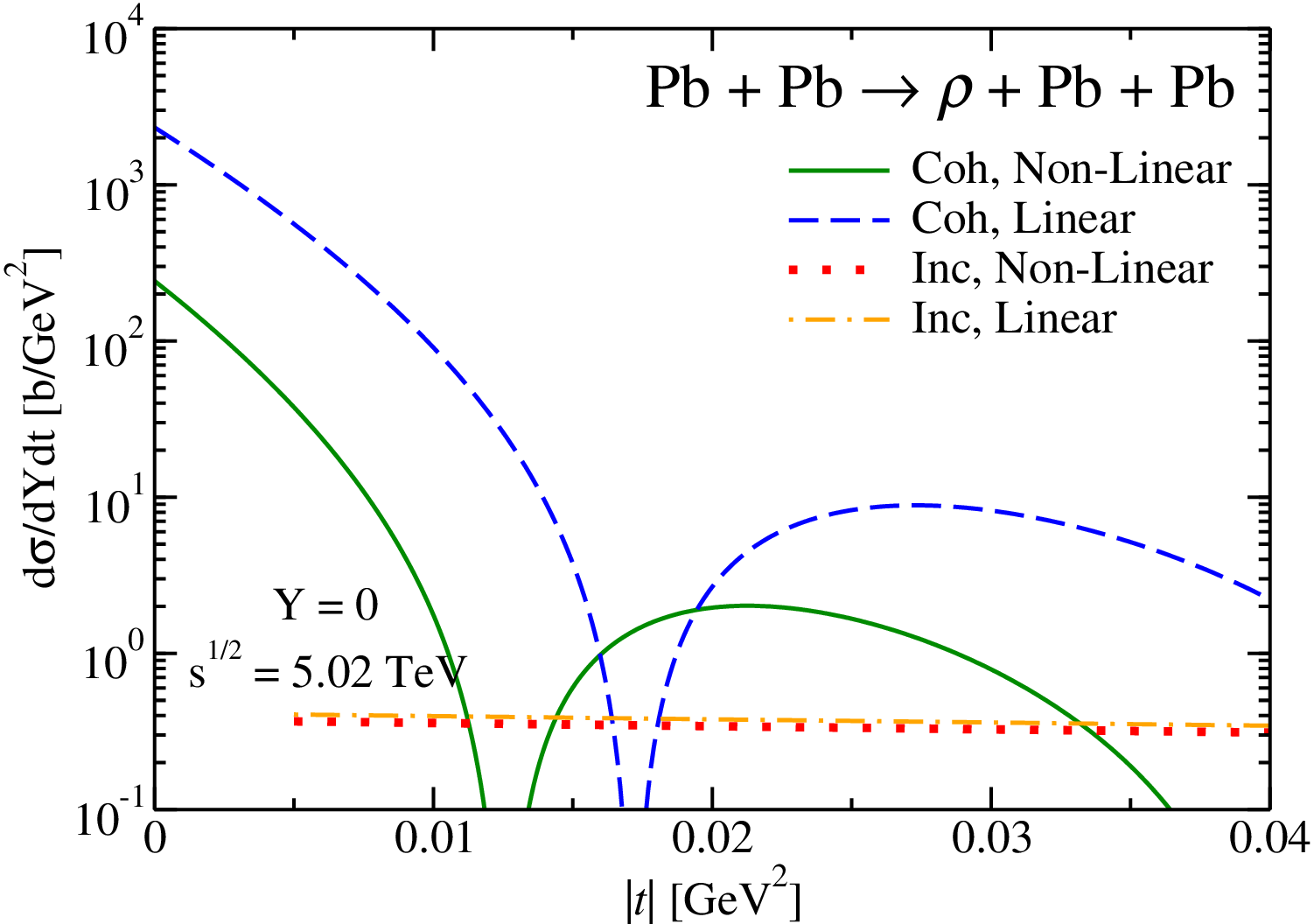}
 \includegraphics[scale=0.28]{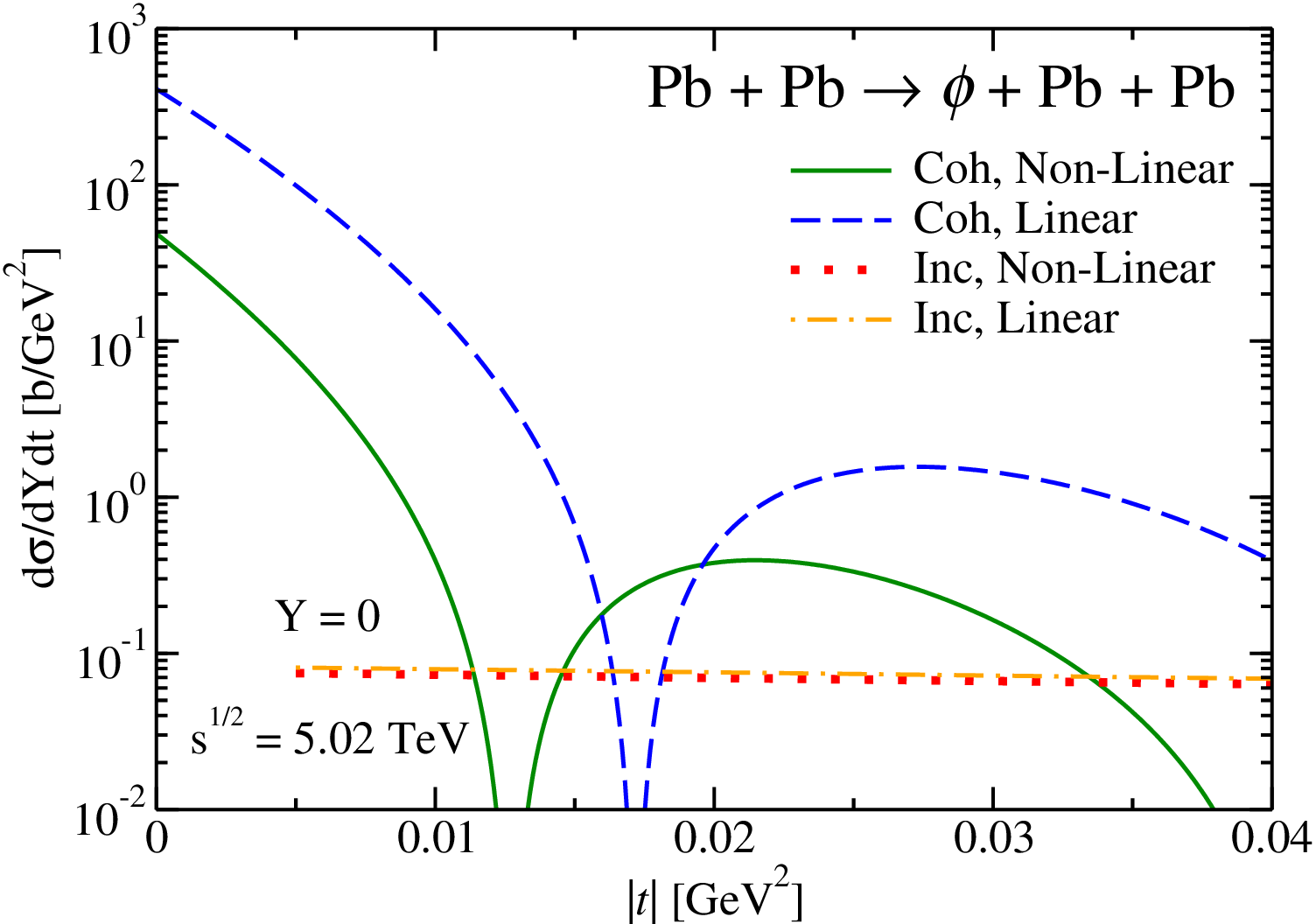}
 \includegraphics[scale=0.28]{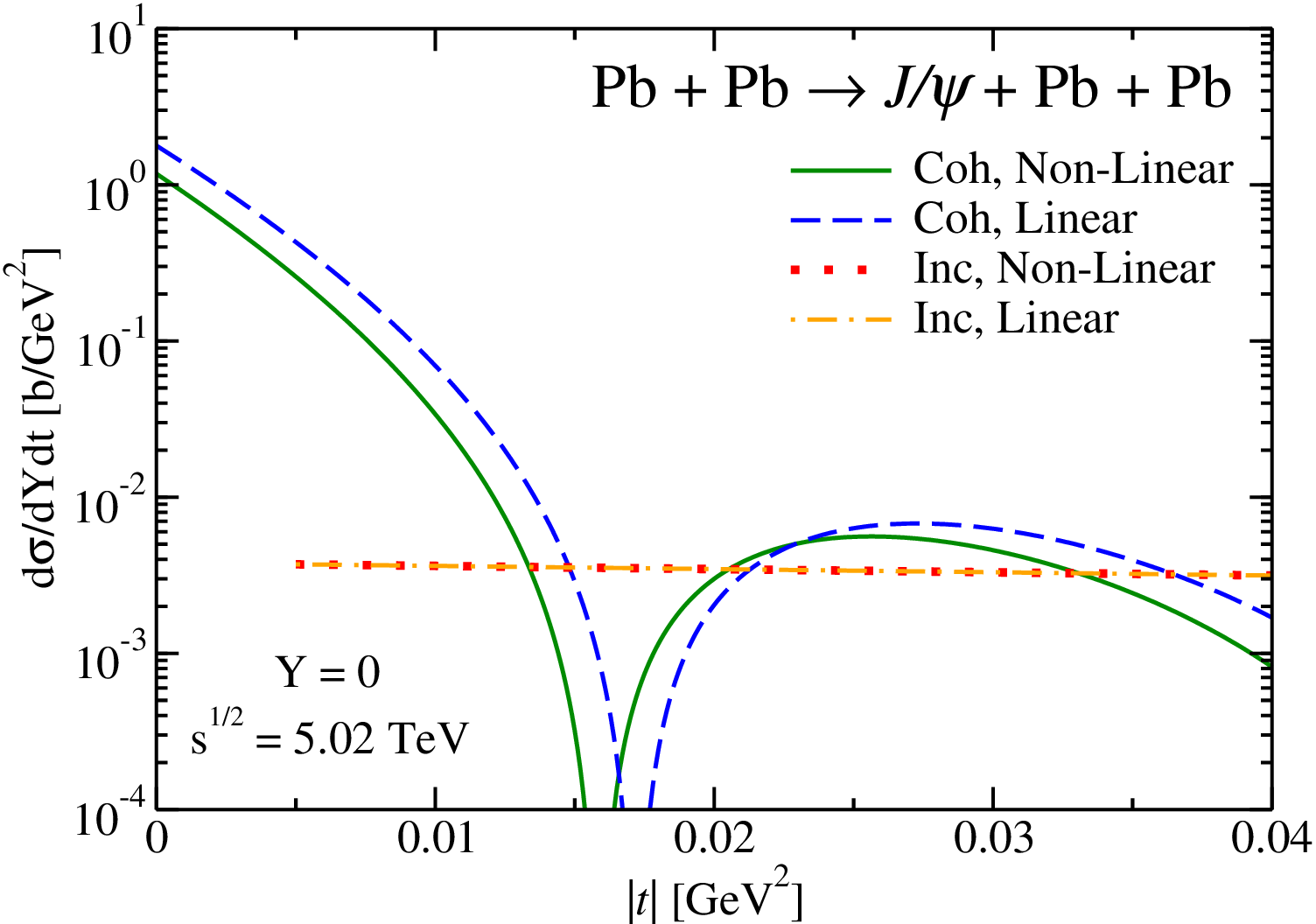}
\caption{Transverse momentum distributions for the exclusive $\rho$, $\phi$ and $J/\Psi$ photoproduction in $PbPb$ collisions at $\sqrt{s} = 5.02$ TeV. The coherent and incoherent contributions are presented separately.}
\label{Fig:dsigdt_AA}
\end{figure}

\begin{figure}[t]
\begin{center}
\scalebox{0.28}{\includegraphics{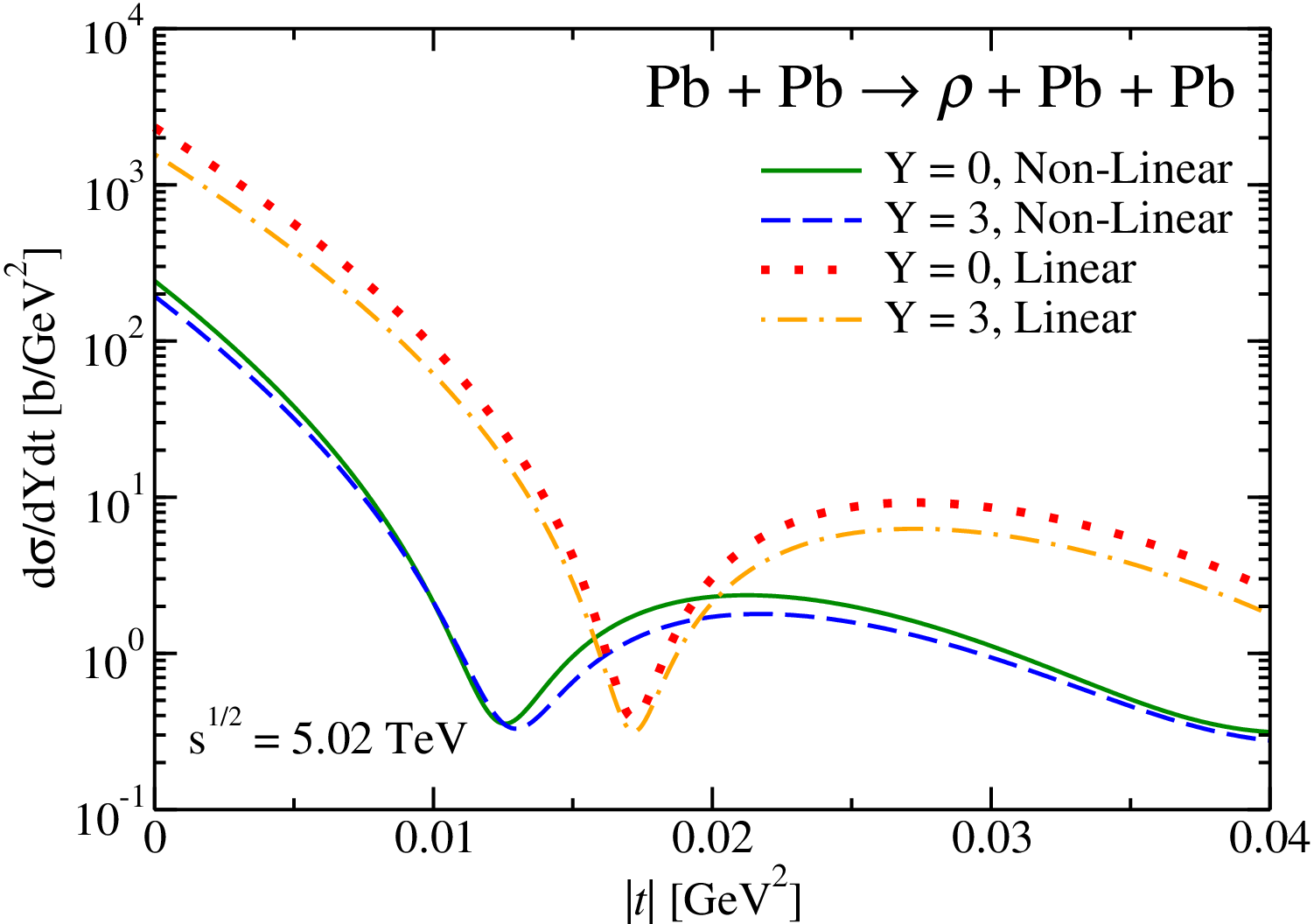}}
\scalebox{0.28}{\includegraphics{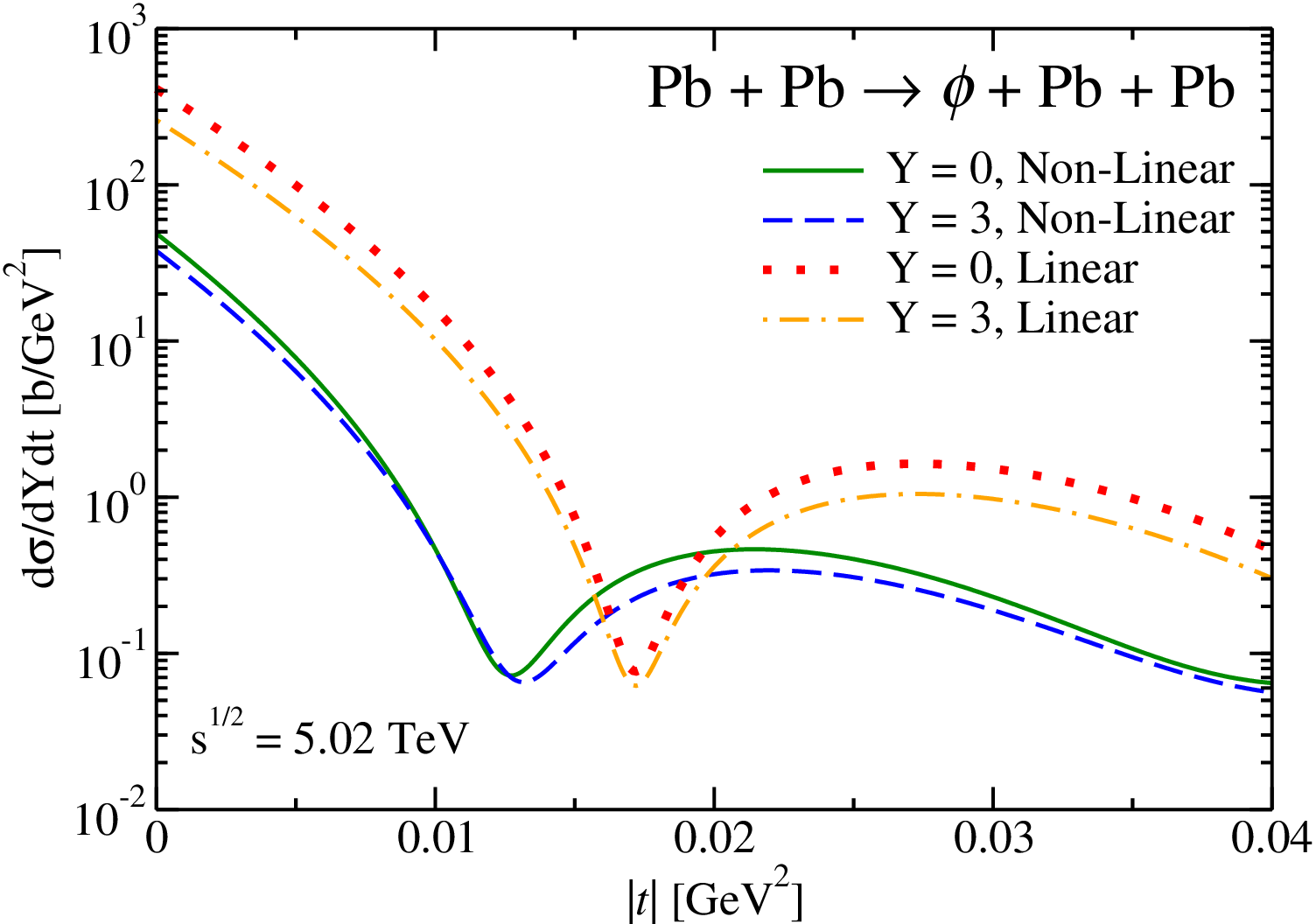}}
\scalebox{0.28}{\includegraphics{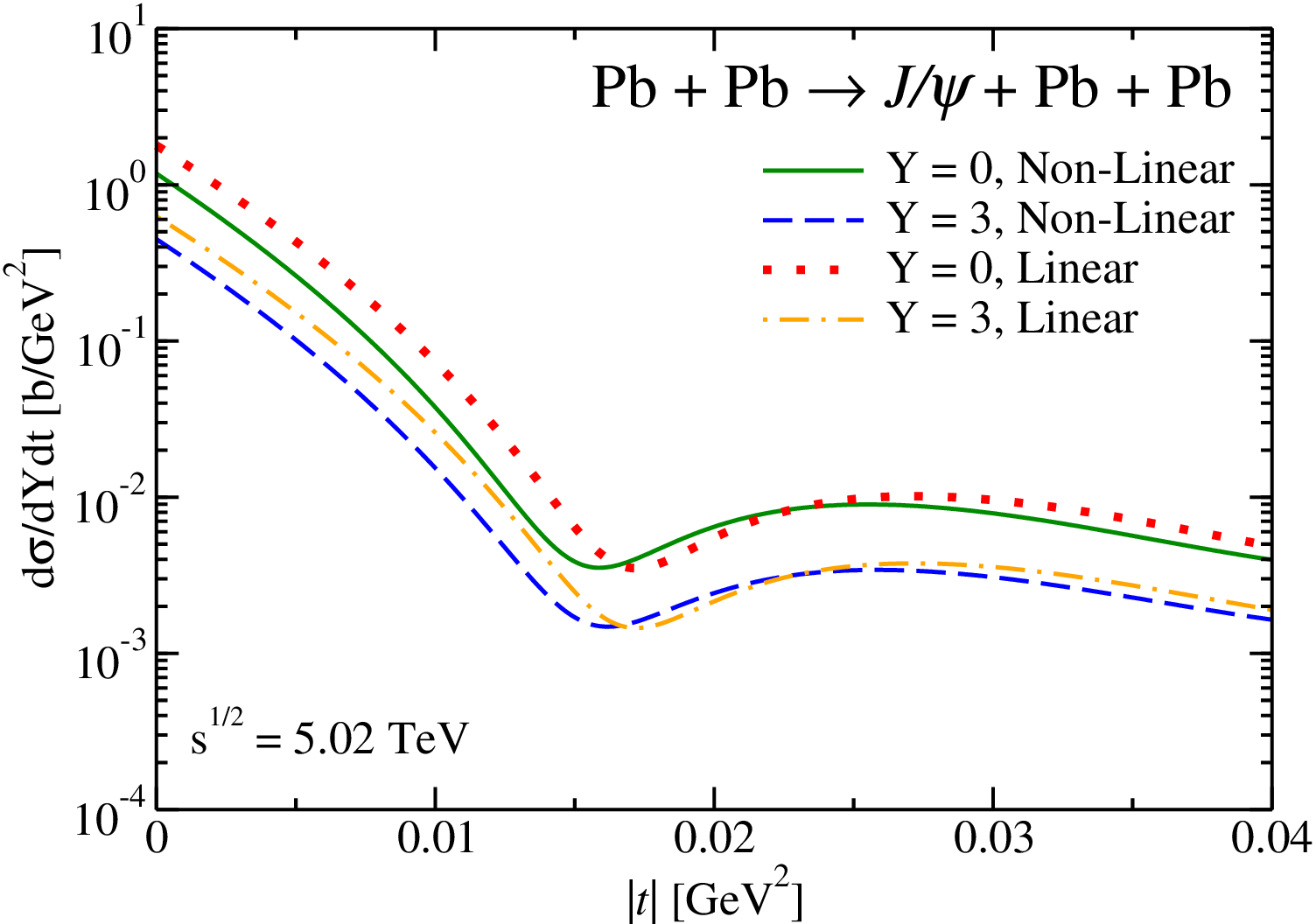}}
\caption{Transverse momentum distributions for the exclusive $\rho$, $\phi$ and $J/\Psi$ photoproduction in $PbPb$ collisions at $\sqrt{s} = 5.02$ TeV. The predictions for the sum of the coherent and incoherent contributions are presented for different values of the vector meson rapidity.}
\label{Fig:dsigdt_AA_sum}
\end{center}
\end{figure}

\section{Results}
\label{res}

In what follows we will present our predictions for  exclusive vector meson photoproduction in 
$pp$ and $PbPb$ collisions at the LHC energies of  Run 2. In particular, we will consider $pp$ collisions at $\sqrt{s} = 13$ TeV and $PbPb$ at 5.02 TeV. Our main focus will be on the transverse momentum distributions, which are expected to be studied considering the higher statistics of  Run 2 
\cite{review_forward}. However, firstly let us analyse the impact of the gluon saturation effects on the rapidity distributions at a fixed value of the momentum transfer $t$. We will estimate  Eq. (\ref{dsigdy}) for $t = t_{min}$, with $t_{min} = - m_N^2 M_V^4/W^4$.
In Fig. \ref{Fig:rap_pp} we present our predictions for the rapidity distributions to be measured in 
$pp$ collisions. We observe that the difference between the linear and non - linear predictions is larger  
for lighter vector mesons, with the gluon saturation effects decreasing the magnitude of the cross sections. In particular, for  exclusive $\rho$ photoproduction, the predictions differ by a factor $\approx 2$ at $Y = 0$. On the hand, for the $J/\Psi$ production,  the predictions are similar. These results are expected, since the gluon saturation effects are predicted to suppress the contribution of the large size dipoles, which are dominant in the $\rho$ case, but contribute less for the  $J/\Psi$ production. Moreover, these results indicate that the analysis of  $\phi$ production is an important probe of the non - linear QCD dynamics. In Fig. \ref{Fig:rap_AA}, we present our predictions for $Pb Pb$ collisions. In this case  the difference between 
the linear and non - linear predictions is larger in comparison to the $pp$ one. The difference is a  factor of the order of 10 at $Y = 0$ for  $\rho$ production, while for $J/\Psi$ production it is 
$\approx$ 2. This result is also  expected, since the saturation scale $Q_s$, which defines the onset of the gluon saturation effects, increases with the atomic mass number ($Q_s^2 \approx A^{1/3}$).  Our results indicate that in  exclusive light vector meson photoproduction in $AA$ collisions we are probing deep in the saturation regime. Moreover, we observe that  gluon saturation effects are non - negligible in the $J/\Psi$ production. As verified in $pp$ collisions, the study of  $\phi$ production can be useful to understand in more detail the QCD dynamics.

Let us now to analyze the impact of the gluon saturation effects on the transverse momentum distributions. Initially, let us consider $pp$ collisions at $\sqrt{s} = 13$ TeV assuming two different fixed values for the vector meson rapidity ($Y = 0$ and 3).  The linear and non - linear predictions for  exclusive $\rho$, $\phi$ and $J/\Psi$ photoproduction are presented in Fig. \ref{Fig:dsigdt_pp}. Our results for $Y = 0$ indicate that the presence of  gluon saturation effects shifts the dip positions to smaller values of the 
transverse momentum, with the shift being larger for lighter mesons, where the contribution of 
these effects is larger. In particular, for the $J/\Psi$ production, the shift is small $\Delta |t| \approx 0.1$ GeV$^2$, while for $\rho$ we have  $\Delta |t| \approx 0.7$ GeV$^2$. Moreover, for the production of light vector mesons, the number of dips in the range $|t| \le 3$ GeV$^2$ is  larger when  gluon saturation effects are present. Another important aspect that can be observed in Fig. \ref{Fig:dsigdt_pp} is that the position of the dip is not modified when we increase the rapidity. However, it is not so pronunced as for central rapidities.

In Fig. \ref{Fig:dsigdt_AA} we present our predictions for $PbPb$ collisions at $\sqrt{s} = 5.02$ TeV. We consider $Y = 0$ and present separately the coherent and incoherent contributions. Similar results are obtained for $Y = 3$.  
In the case of the incoherent predictions we only present predictions for $|t| \ge 0.005$ GeV$^2$, since the model proposed in Ref. \cite{Lappi_inc} and used in our calculations fails to describe the vanishing of the incoherent cross section as $|t| \rightarrow 0$. As expected, we find that the coherent cross section clearly exhibits the typical diffractive pattern. Moreover, the coherent processes are characterized by a sharp forward diffraction peak and the incoherent one by a weak $t$ - dependence. We have verified that the incoherent processes dominate at large - $|t|$  and the coherent ones at small values of the momentum transfer. 
This is  expected, since  increasing  the momentum kick given to the nucleus the probability that  
it  breaks up becomes larger. Additionally,  the presence of gluon saturation effects strongly decreases the magnitude of the coherent cross sections, in particular for lighter vector mesons, and implies a shift in the position of the dip to  smaller values of $t$. In the case of the incoherent contribution, we have that the linear and non-linear predictions are similar.

Our results indicate that incoherent processes dominate at large - $|t|$  and the coherent ones at small values of the momentum transfer.  Therefore, one can expect that the analysis of the $t$  dependence can be useful to separate coherent and incoherent interactions. However, as discussed in detail in Refs. \cite{Caldwell,Toll}, the experimental separation of these processes is still  a challenge.  An alternative is the detection of the fragments of the nuclear breakup produced in the incoherent processes. e.g. the detection of emitted neutrons by zero - degree calorimeters. Considering that this separation  is not yet possible, 
in Fig. \ref{Fig:dsigdt_AA_sum} we present our predictions for the sum of the coherent and incoherent contributions. We observe  that the incoherent contribution partially fills the dip in the tranverse momentum distribution. However, it is still  present, with its position being affected by  gluon saturation effects.

\section{Conclusions}
\label{conc}
The study of  exclusive vector meson photoproduction in hadronic collisions  is strongly motivated by the expectation that this process  may allow us to  probe  the QCD dynamics at high energies,  driven by the gluon content of the target (proton or nucleus) which is strongly sensitive to non-linear effects (parton saturation).
Our goal in this paper was to extend  and complement previous studies about  exclusive vector meson 
photoproduction in $pp$ and $PbPb$ collisions, presenting  the color dipole 
predictions for the transverse momentum distributions taking into account gluon saturation effects 
in the QCD dynamics. In particular, we have used an approach that reproduces well the available HERA data on 
vector meson photo and electroproduction, including data on the  $t$-distributions, as well as the Run 1 
LHC data on  vector meson photoproduction. We presented predictions for the $t$ - spectrum of the exclusive $\rho$, $\phi$ and $J/\Psi$ photoproduction in $pp$ and $PbPb$ collisions, which could be compared with future experimental LHC data. In order to estimate the impact of the gluon saturation effects, we also have presented a comparison with the predictions obtained disregarding these effects. Our results demonstrate that gluon saturation effects reduce 
the magnitude of the cross sections, with the reduction being larger for lighter vector mesons. 
Moreover,  the gluon saturation effects change the  positions of  the dips with respect to the linear regime, shifting the dips to smaller values of the transverse momentum. Finally, our results indicate 
that dips predicted by the coherent contribution in $PbPb$ collisions should be visible, independently of the fact that this contribution could not been easily experimentally separated.  These results are robust 
predictions of the saturation physics, which can be used to investigate non-linear QCD dynamics in the kinematical range of the Run 2 of the LHC.


\begin{acknowledgements}
 This work was  partially financed by the Brazilian funding agencies CNPq, CAPES, FAPESP and FAPERGS.
\end{acknowledgements}

\end{document}